\documentclass[showpacs,preprintnumbers,amsmath,amstex,amssymb,letterpaper]{jpconf}
\usepackage{graphicx}
\usepackage{enumerate}
\usepackage{bm}
\usepackage{slashed}

\usepackage{amssymb}
\usepackage{amsmath}
\usepackage{epsfig}
\newcommand{\be}{\begin{equation}}
\newcommand{\ee}{\end{equation}}
\newcommand\beq{\begin{eqnarray}}
\newcommand\eeq{\end{eqnarray}} 
\newcommand\eqn[1]{\label{eq:#1}} 
\newcommand\eq[1]{eq.~(\ref{eq:#1})}

\newcommand{\vev}[1]{\langle #1 \rangle}

\newcommand{\bfp}{{\mathbf p}}

\newcommand{\CM}{{\cal M}}
\newcommand{\CN}{{\cal N}}
\newcommand{\CP}{{\cal P}}

\newcommand{\CL}{{\cal L}}

\newcommand\expect[3]{\langle #1|#2|#3\rangle}

\newcommand{\mybar}[1]%
        {\kern 0.6pt\overline{\kern -0.6pt#1\kern -0.6pt}\kern 0.6pt}
\begin{document}


\title{Elucidating the sign problem through noise distributions}

\author{Amy N. Nicholson$^{1}$, Dorota Grabowska$^2$ and David B. Kaplan$^2$}
  
\address{$^1$ Maryland Center for Fundamental Physics, Department of Physics,
University of Maryland, College Park MD 20742-4111, USA}

\address{$^2$Institute for Nuclear Theory, University of Washington, Seattle, WA 98195-1550, USA}

 \ead{amynn@umd.edu, grabow@uw.edu, dbkaplan@uw.edu}


 
 \begin{abstract}
Due to the presence of light pions in the theory, lattice QCD at finite densities suffers from issues with noise in both grand canonical and canonical formulations. We study two different formulations of the Nambu-Jona-Lasinio model reduced to 2+1 dimensions at large $N$, where $N$ is the number of flavors. At finite chemical potential one formulation has a severe sign problem and a fermion correlator which displays a broad probability distribution with small mean. In the other we find no sign problem and a distribution amenable to the cumulant expansion techniques developed in Ref.~\cite{Lee:2011sm,Endres:2011jm,Endres:2011mm}.

\end{abstract}


\section{Introduction}
While lattice QCD is currently the only tool for studying QCD from first principles in the non-perturbative regime, the study of QCD at finite densities on the lattice has been greatly hindered by two related problems. In the first, lattice QCD at fixed baryon chemical potential suffers from what is known as a sign problem, which arises due to the complexity of the fermion determinant one wishes to use as a probability measure for Monte Carlo calculations. A possible solution is to employ phase reweighting, in which the phase of the determinant is absorbed into the observable and the magnitude of the determinant is used as a probability measure; however, the expectation value of the phase using the same ensemble must also be calculated in order to reconstruct the original observable. One finds that for $\mu$ above some critical value, this expectation value vanishes exponentially and becomes swamped by statistical noise. The second problem is encountered in a canonical formulation, in which a fixed number of quark sources and sinks are separated by a long Euclidean time to project out the ground state. Again, the signal-to-noise ratio is found to vanish exponentially with time.

A notable property of these two problems is that they are both related to the same dynamical property of QCD; namely, they both arise due to the presence of light pions and thus are intimately related to the phenomenon of chiral symmetry breaking. In the case of finite chemical potential, this can be seen by noting that for two degenerate flavors of quark the magnitude of the fermion determinant corresponds to chemical potential for isospin. Thus, at low temperatures for $\mu \ge m_{\pi}/2$ pion condensation would tend to occur, and the role of the phase must be to cancel this pion condensation \cite{Splittorff:2006fu,Splittorff:2007ck}. 

For the other case where finite baryon density is achieved using correlation functions of quark propagators, the role of the pion in the signal-to-noise problem may be noted by considering the variance of an operator used to create nucleons. While at large Euclidean times the correlator for three appropriately contracted quark propagators projects out the ground state of the nucleon, $M_N$, the variance will be composed of three quark and three anti-quark propagators, projecting out the pion mass at large times \cite{Parisi:1983ae,Lepage:1989hd}. The signal-to-noise ratio will thus be exponentially vanishing with the factor $(M_N-\frac{3}{2}m_{\pi})\tau$, where $\tau$ is the Euclidean time, for late Euclidean times.

Both of the noise problems presented above can be shown to arise from probability distributions which are broad, symmetric, and centered about zero. For the nucleon correlator, one may see this by looking at higher moments of the distribution function. Even moments consist of equal numbers of quark and antiquark propagators which may pair up to form pions, while odd moments have three unpaired quark propagators which will form a baryon \cite{Savage:2010}. Thus, because odd moments will always contain a baryon mass in the exponential, they will be suppressed relative to the even moments, giving a symmetric distribution with exponentially small mean. A similar argument may be made for the expectation value of the phase at finite chemical potential, where all odd moments are equivalent to the mean, and therefore exponentially small for $\mu \geq m_{\pi}/2$, while all even moments give exactly $1$ (for a discussion of the distribution of the phase, see \cite{Lombardo:2009aw}). 

In \cite{Lee:2011sm,Endres:2011jm,Endres:2011mm} it was shown that another type of noise problem exists in which the distribution displays a long tail, resulting in a sample mean which may be far from the true mean with deceptively small error bars. This problem is often referred to as an overlap problem, and has been known to occur for observables when reweighting is performed. This was discussed in a canonical formulation using lattice calculations of unitary fermions, a strongly-interacting nonrelativistic system in which the fermion determinant is real and positive for even numbers of fermion flavors. The probability distributions for fermion correlators were shown to be nearly log-normal, and a possible solution to the overlap problem was postulated in the form of a truncated cumulant expansion for the correlator,
\beq
\ln \langle \mathcal{O}(\phi) \rangle = \sum_{n=1}^{N_{\kappa}} \frac{\kappa_n(\ln \mathcal{O})}{n!}\ ,
\eeq
where $\mathcal{O}(\phi)$ is some operator of interest on a background field $\phi$, $\kappa_n(\ln \mathcal{O})$ is the $n$th cumulant of the logarithm of the operator, and $N_{\kappa}$ is some suitable truncation order, determined by examining the convergence of the series. For a distribution that is sufficiently close to log-normal, this series may converge at a low enough order to be useful in practical applications. Recently it has also been reported that nearly log-normal distributions are also seen in lattice QCD calculations of correlation functions for intermediate times \cite{DeGrand:2012ik}. 

\section{Two formulations for a large-$N$ NJL model in $2+1$-dimensions}
To understand the origin of these two types of noise problem in QCD and to further explore the link between signal-to-noise and the presence of light pions, we turn now to a QCD-like model which displays chiral symmetry breaking without the added complication of confinement. For more details behind these calculations, see \cite{Grabowska:2012ik}. The model we will consider is the NJL model in $2+1$-dimensions for large number of fermion flavors, $N$,
\beq
\CL = N\left(\mybar\psi_a(\slashed{\partial}-m)\psi_a- \frac{C}{2}\left[(\mybar\psi_a\psi_a)^2 + (\mybar\psi_a i\gamma_5\psi_a)^2\right]\right)\ ,
\eqn{NJL}\eeq
where $a,b\ldots$ are a flavor index  summed over $1,\ldots,N$, indices, $i,j\ldots$ are summed over 3D coordinates $1,2,3$, while Greek indices $\mu,\nu...$ are summed over 4D coordinates $1,\ldots,4$. The gamma matrices are the usual $4\times 4$ matrices used in 4D, and so the Lagrangian represents $2N$ flavors of 3D Dirac fermions.  In the limit $m\to 0$ this theory has a $U(1)$ chiral symmetry in 4D, which becomes a flavor symmetry in 3D; this ``chiral'' symmetry is spontaneously broken as in $3+1$-dimensions, giving rise to a Goldstone boson.

To study this model numerically one may introduce auxiliary fields to generate a four-fermion interaction, which can then be integrated over using Monte Carlo methods. We find two physically equivalent methods for doing so. The first, which we will call the $\sigma/\pi$ formulation, is the conventional method which introduces scalar fields that are singlet under the $SU(2N)$ flavor symmetry, 
\beq
\CL =N\left( \frac{1}{2C}\left(\sigma^2+\pi^2\right) +\mybar\psi_a \left[\slashed{\partial}-m+\sigma + i\pi\gamma_5\right]\psi_a\right)\ .
\eqn{SP}\eeq
The second formulation, which we will call the $A/V$ formulation, follows if one performs a Fierz rearrangement of the four-fermion interaction in \eq{NJL} before introducing auxiliary fields. This results in the introduction of $N\times N$ matrix valued vector and axial vector auxiliary fields $V$ and $A$, and the equivalent theory
\beq
\CL =N\left( \frac{1}{C}\, \Tr\left(V_\mu V_\mu+A_\mu A_\mu \right) +\mybar\psi_a \left[\slashed{\partial}-m+i\slashed{V} + \slashed{A}\gamma_5\right]_{ab}\psi_b\right)\ .
\eqn{VA}\eeq

To determine whether these two formulations will cause a Monte Carlo sign problem at non-zero chemical potential, we should investigate the positivity of the fermion determinants. For the $\sigma/\pi$ formulation, it is possible to define a real symmetric charge conjugation matrix $C$ satisfying $C^2=1$, $C\gamma_\mu C=\gamma_i^*$ for $i = 1,2,3$ and $C\gamma_5 C = -\gamma_5^*$ . Then the fermion operator for a single flavor in the grand canonical formulation satisfies $D^* = CDC$, where $D= (\slashed \partial -m + \sigma + i\pi\gamma_5 + \mu \gamma_1)$, and complex eigenvalues of $D$ must come in conjugate pairs.  Thus $(\det D)^N$  is real, and positive for even $N$. This implies that there is no sign problem at finite density \cite{Hands:1995jq}.

In the $A/V$ formulation, the fermion matrix at finite chemical potential is given by $D(\mu) = \left(\slashed{\partial}-m+i\slashed{V} + \slashed{A}\gamma_5+\mu\gamma_1\right)$, which is similar in structure to the QCD Dirac matrix with nonzero $\mu$, and its determinant is similarly complex.  In fact, as in QCD, the magnitude of the fermion determinant for two degenerate families, $\vert \det D(\mu)\vert^2$, corresponds to isospin chemical potential, so that for $\mu\geq m_{\pi}/2$, the phase is responsible for eliminating pion condensation in the ground state.

Chiral symmetry breaking is simple to see in the $\sigma/\pi$ formulation, where the large $N$ expansion is equivalent to the semiclassical expansion. Upon integrating out the fermion fields we have the following effective action:
\beq
S[\sigma,\pi] =N \int {\rm d}^3x  \left[ \frac{ 1}{2C } (\sigma(x)^2+\pi(x)^2) -\Tr\ln(\slashed{\partial}-m+\sigma(x)+i\pi(x)\gamma_5)\right] \ .
\eqn{ssp}\eeq
Performing a mean field calculation leads to (using dimensional regularization and the MS subtraction scheme), 
\beq
\vev{\sigma} = \frac{f}{2}\left[1+\sqrt{1+4m/f} + 2m/f\right]\ , \qquad M \equiv \vev{\sigma}-m = \frac{f}{2}\left[1+\sqrt{1+4m/f} \right] \ ,
\label{eq:sigmaM}
\eeq 
where we have defined $\langle \sigma \rangle = f$ to be the chiral symmetry breaking minimum when $m=0$, and $M$ is the constituent fermion mass. We may also find the $\sigma$ and $\pi$ dispersion relations $D_\sigma(k), D_\pi(k)$ by expanding the effective action to second order about our mean field solution.

\begin{figure}[t]
\includegraphics[width=2.4in]{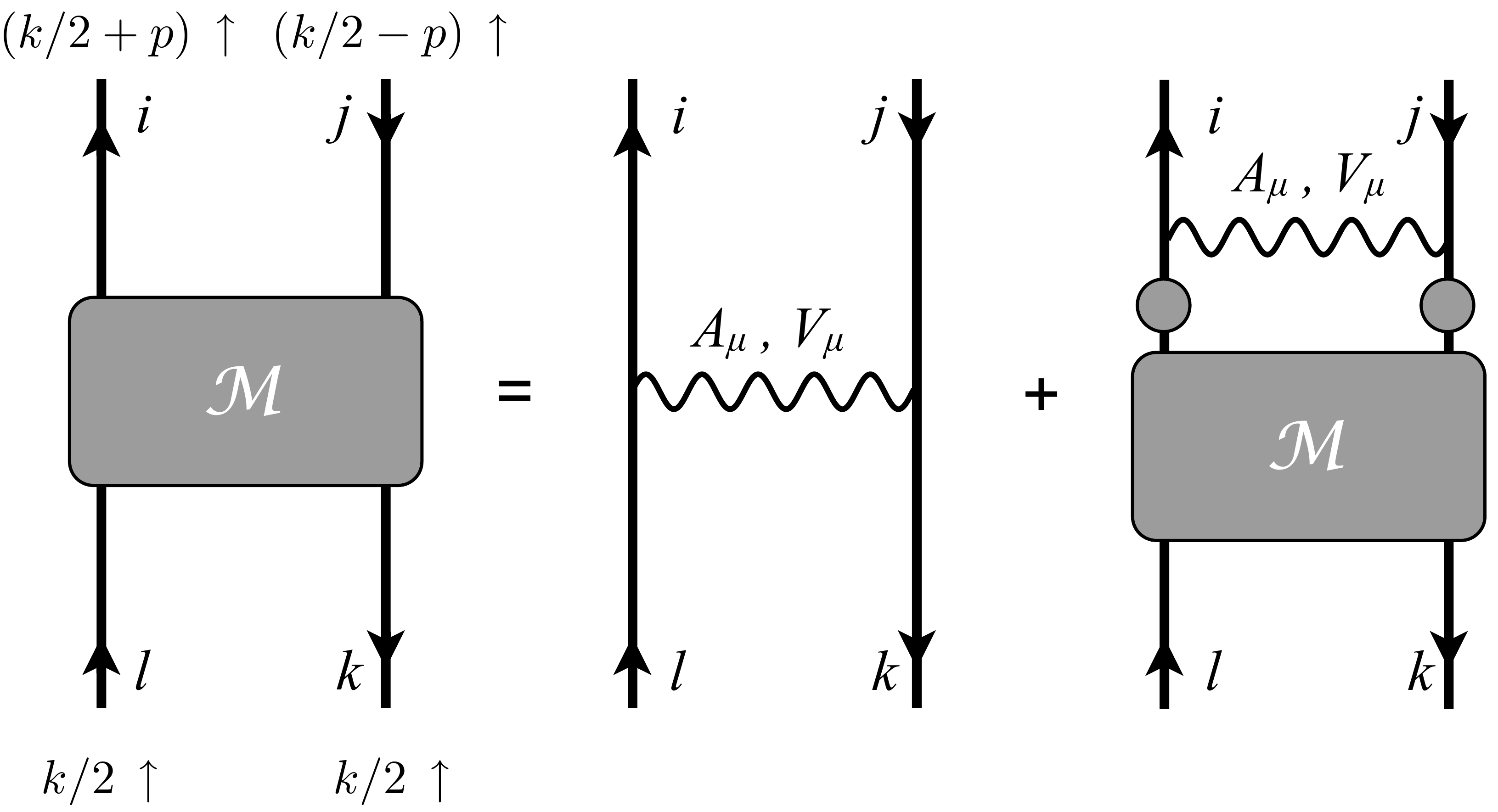}\hspace{2pc}%
\centering{\begin{minipage}[b]{20pc}\caption{\label{fig:fourpoint}\it Graphical representation of the integral equation for the four-point correlator in the $A/V$ formulation for an incoming fermion-antifermion pair of one flavor and an outgoing pair of another. Dirac indices are labeled.}
\end{minipage}}
\end{figure}

In the $A/V$ formulation we cannot use mean field theory; instead, to leading order in $1/N$ we may find the fermion propagator by solving the Schwinger-Dyson equation and finding the nonzero fermion mass, Eq.~\ref{eq:sigmaM}. Furthermore, one may derive the $\sigma$ and $\pi$ meson propagators by solving the matrix equation shown in Fig.~\ref{fig:fourpoint}. One finds $\CM_{ij;kl}(k)  = -\left(  \delta_{ij}\delta_{kl}/ D_\sigma(k) +  (i\gamma_5)_{ij}(i\gamma_5)_{kl}/D_\pi(k) \right)$. Thus we see that an interaction via $t$-channel exchange of $A_\mu$ and $V_\mu$ mesons is exactly equivalent to a single meson in the $s$ or $u$-channel in the $\sigma/\pi$ formulation, corresponding to a valence fermion/antifermion pair or two valence fermions or antifermions, respectively.

\section{Probability distributions for fermion correlators}
We now proceed to calculate the probability distribution for a fermion propagator in both formulations. If $X[\phi]$ is a functional of a stochastic field $\phi$ corresponding to an observable, we define the normalized probability density function for $X$ to be the path integral $\CP(x) =\CN \int [{\rm d}\phi]\, e^{-S[\phi]} \,\delta(X[\phi]-x)$, where we assume $S[\phi]$ is real. This quantity is difficult to analyze field theoretically due to the singular nature of the delta function, so we perform a Fourier transform and work with the characteristic function,
\beq
\Phi(s, \bar s) = e^{-W(s, \bar s)} =  \CN\int [{\rm d}\phi]\, e^{-S[\phi] + i (s X[\phi]+\bar s \bar X[\phi])} \ ; \qquad W(s, \bar s) = -\sum_{m,n=1}^\infty \frac{(i s)^m(i \bar s)^n}{m!\,n!}\, \kappa_{m,n}\ ,
\eqn{compphi}\eeq
where we have generalized the cumulant generating function $W(s,\bar s ) =- \ln \Phi(s, \bar s)$ for a complex observable through a double expansion in both $s$ and $\bar s$. From this formulation we see that we may compute cumulants using connected Feynman diagrams of the modified action $S[\phi]-i (s X[\phi]+\bar s \bar X[\phi])$.

For our observables we will use
\begin{subequations}
\begin{align}
 Y_\Gamma&=\ln\left[\frac{1}{V}\expect{\bfp=0,t=T/2\,}{\Tr \Gamma\left(\frac{1}{\slashed{\partial} -m+\sigma + i \pi\gamma_5}\right)}{\bfp=0,t=-T/2}\right]\quad &\sigma/\pi\text{ formulation,}\eqn{corrsp}\\
 X_\Gamma&=\frac{1}{V}\expect{\bfp=0,t=T/2\,}{\Tr \Gamma\left(\frac{1}{\slashed{\partial} -m+i\slashed{V} + \slashed{A}\gamma_5}\right)}{\bfp=0,t=- T/2}\quad &A/V\text{ formulation,}\eqn{corrAV}
 \end{align}
 \end{subequations}
 where $\Gamma$ is some Dirac matrix of our choosing, and we consider the logarithm of the propagator in the $\sigma/\pi$ case to simplify the calculation later on. Measuring the expectation value of this correlator is a procedure for determining the mass $m_f$ of the lightest fermion state allowed for a given $\Gamma$ using $\lim_{T\to\infty} \ -\frac{1}{T} \ln \vev{X_\Gamma} = \lim_{T\to\infty} \ -\frac{1}{T} \ln \vev{e^{Y_\Gamma}} =m_f$.

The calculation of the variance for the $A/V$ case involves attaching fermion propagators at zero spatial momentum to the legs in the first diagram in Fig.~\ref{fig:fourpoint}. For $\Gamma=1$ and near the chiral limit the pion is the lightest state which can propagate through such a graph, so that for late Euclidean time we find, $\kappa_{1,1} \simeq \frac{2\pi }{ m_\pi f V N } e^{-m_\pi T}$. Higher cumulants can be computed using the equivalent $\sigma/\pi$ diagrams. Generically, a graph contributing to $\kappa_{m,n}$ scales as $N^{m+n-1}$, since we need a minimum of $(m+n-1)$ mesons to make a connected graph. Furthermore, it is straightforward to verify that the minimum mass state that can possibly propagate in a graph for $\kappa_{m,n}$ with $m\ge n$ consists of $(m-n)$ fermions with mass $M$  and $n$ pions.  Therefore we expect these cumulant to scale as
\beq
\kappa_{m,n}\sim \frac{e^{-(m-n)MT}e^{-n m_\pi T}}{N^{m+n-1}}\qquad\qquad\qquad (m\ge n)\ .
\eeq
The above scaling implies that the distribution for the real part of the fermion propagator near the chiral limit becomes highly symmetric about zero at late time.  That is because odd moments (for which $(m-n)\ne 0$) are seen to fall off much more quickly than even moments.  This is consistent with the Lepage-Savage picture for baryon propagator distributions in QCD.

 For the $\sigma/\pi$ formulation the cumulants are given by the connected graphs derived from the action $S_Y = N S[\sigma,\pi]-i s Y_\Gamma[\sigma,\pi]$, where $S[\sigma,\pi]$ is given in \eq{ssp}. At leading order in a $1/N$ expansion, $\kappa_n$ will be given by the sum of all tree level diagrams composed of $n$ external lines and any number of vertices which arise from internal fermion loops. Given our definition of $Y_{\Gamma}$, we find that all external propagators are spatially homogeneous, and, for very large Euclidean time, temporally homogeneous up to edge effects near the source and sink. Because we only consider tree graphs, all internal meson propagators will similarly be spatially and temporally homogeneous; thus, we may once again use mean field theory. Solving for the mean field leads to the following cumulants for the logarithm of the fermion correlator:
\beq
\kappa_1 &=&\ln z-M T\ ,\qquad
\kappa_{n\ge 2} = -\frac{(2(n-2))!}{(n-2)!}\, (2 M-f) T\, \zeta^{n-1} \ , \qquad \zeta \simeq \frac{\pi }{N V (2M-f)^2}\ \,
\eqn{kappasp} \eeq
where $z$ gives the overlap of our operator with the ground state, and we have approximated the integral over Euclidean time by a step function for $-T/2\leq \tau \leq T/2$. Note in the limit $N,T,V \to \infty , \ \frac{T}{VN} = $ finite, the cumulants $\kappa_n$ vanish for $n\ge 3$ and $Y_\Gamma$ assumes a normal distribution, giving a log-normal distribution for the correlator. With the variance for the log of the correlator growing linearly with time, the distribution for the correlator will eventually become heavy-tailed, with a skewness growing exponentially with $\sigma$, though such long times are unlikely to be reached in practical numerical simulations.

\section{Discussion}

We found that in the QCD-like ``$A/V$" formulation of the NJL model, the fermion determinant was complex and a Splittorff-Verbaarschot argument \cite{Splittorff:2006fu, Splittorff:2007ck} could be made to show that the phase of the fermion determinant had to fluctuate wildly for $\mu>m_\pi/2$.  When looking at fermion correlators, the distribution evolved to be symmetric about an exponentially small mean relative to its width, implying a severe signal/noise ratio when sampling the correlator using Monte Carlo methods. In contrast the ``$\sigma/\pi$" formulation with even $N$ has no sign problem at nonzero $\mu$, and the correlator distribution was, in a certain limit, log-normal and heavy-tailed.  This overlap problem would pose challenges to Monte Carlo sampling if the tail became too long, but this sort of problem seems to be less severe than the exponential fall-off of signal/noise seen in the $A/V$ formulation as seen with the cumulant expansion analysis of Ref.~\cite{Endres:2011jm,Endres:2011er, Endres:2012cw}.

Our analysis should make it clear that the sign problem encountered in QCD at nonzero chemical potential is not a fermion problem, but instead a consequence of interactions.  In particular, if the particles being studied can exist in a tightly bound state of valence fermions, there is going to be a sign problem.  We believe that inventing a way to introduce the pion into QCD as a fundamental field could be an important step toward solving the QCD sign problem and beginning to study the properties of ordinary and dense matter from first principles.

\ack
This work was supported in part by U.S.\ DOE grants No.\ DE-FG02-00ER41132 andNo.\ DE-FG02-93ER-40762.

\section*{References}
\bibliographystyle{iopart-num}
\bibliography{NJLref}

\end{document}